\documentclass[twocolumn]{aastex62}

\usepackage{amsmath}
\usepackage[T1]{fontenc}
\usepackage{apjfonts} 

\newcommand{\code}[1]{\texttt{#1}}
\newcommand{\mesa}{\code{MESA}}
\newcommand{\MESA}{\mesa}
\newcommand{\abs}[1]{{\left|{#1}\right|}}
\newcommand{\BV}{Brunt-V\"{a}is\"{a}l\"{a}}

\newcommand{\paperone}{Paper~I}  

\newlength{\apjcolwidth}
\newlength{\figwidth}
\newlength{\doublewide}
\begin{document}
\title{Polluted White Dwarfs: Mixing Regions and Diffusion Timescales}

\author[0000-0002-4791-6724]{Evan B. Bauer}
\affiliation{Department of Physics, University of California, Santa Barbara, CA 93106, USA}
\correspondingauthor{Evan B. Bauer}
\email{ebauer@physics.ucsb.edu}

\author{Lars Bildsten}
\affiliation{Department of Physics, University of California, Santa Barbara, CA 93106, USA}
\affiliation{Kavli Institute for Theoretical Physics, University of California, Santa Barbara, CA 93106, USA}

\begin{abstract}
Many isolated white dwarfs (WDs) show spectral evidence of atmospheric metal
pollution. Since heavy element sedimentation timescales are short,
this most likely indicates ongoing accretion. Accreted metals
encounter a variety of mixing processes at the WD surface: convection,
gravitational sedimentation, overshoot, and thermohaline instability.
We present \mesa\ WD models that explore each of these processes and
their implications for inferred accretion rates. We
provide diffusion timescales for many individual metals, and we
quantify the regimes in which thermohaline mixing dominates over
gravitational sedimentation in setting the effective settling rate of
the heavy elements. We build upon and confirm earlier work
finding that accretion rates as high as $10^{13} \, \rm g \, s^{-1}$
are needed to explain observed pollution in DA WDs for
$T_{\rm eff} > 15,000 \, \rm K$, and we provide
tabulated results from our models that enable accretion rate
inferences from observations of polluted DA WDs. If these rates are
representative of young WDs, we estimate that the total mass of planetesimal
material accreted over a WD lifetime may be as high as $10^{28} \, \rm
g$, though this estimate is susceptible to potential selection biases
and uncertainties about the nature of disk processes that supply
accretion to the WD surface. We also find
that polluted DB WDs experience much less thermohaline mixing than DA
WDs, and we do not expect thermohaline instability to be
active for polluted DB WDs with $T_{\rm eff} < 18,000 \, \rm K$.
\end{abstract}


\section{Introduction}

A large fraction ($25\%$--$50\%$) of isolated white dwarf (WD)
atmospheres show signatures of polluting metals
\citep{Zuckerman03,Koester14}. Heavy element sedimentation timescales
are short \citep{Schatzman45,Schatzman48}, and 
this implies recent or ongoing accretion of
observed heavy elements \citep{Vauclair79,Koester14}. Polluted WD
spectra are often accompanied by infrared emission from a dust
disk \citep{Koester97,Farihi09,Girven12,Farihi16}, and the predominant
model for the origin of this dust is debris from disrupted
planetesimals \citep{Jura03,Jura14,Vanderburg15}. Models for WD
surface mixing allow inferences of the composition of these
planetesimals and the rates at which WDs accrete this material
\citep{Koester06,Koester09,Dufour10,Dufour12,Koester11,Farihi13,Raddi15}.

While many have relied on
elemental sedimentation timescales to make inferences about polluted
WD accretion, recent work has revealed that thermohaline instability
is active and substantially modifies the inferred accretion rate
$\dot M_{\rm acc}$ \citep{Deal13,Wachlin17}. Our work in
\citet[][\paperone]{Bauer18} greatly expanded the range of
$T_{\rm eff}$ explored for polluted WD models accounting for
thermohaline mixing.
In \paperone, we found that some DA WDs must accrete bulk earth
composition at rates as large as 
$\dot M_{\rm acc} = 10^{13} \, \rm g \, s^{-1}$ for our models to
match observed surface metal abundances.

In this work, we build on the results of \paperone\ with further
examination of the surface mixing processes relevant for heavy element
pollution.
We construct models that include these processes using the stellar
evolution code \mesa\ \citep{Paxton11,Paxton13,Paxton15,Paxton18}. We
use \mesa\ version 10398 unless otherwise stated.
For models that include thermohaline mixing, we
use \mesa\ version 11191 (see Section~\ref{S.Thermohaline} for a
description of the relevant changes to the code that motivate using
this version). \mesa\ inlists and other input files necessary to
reproduce our \mesa\ models are available online at
\url{https://doi.org/10.5281/zenodo.2541235}.
In Section~\ref{S.convection}, we describe the hydrogen-dominated
surface convection zones that metals first encounter when
they accrete onto DA WDs.
In Section~\ref{S.diffusion}, we quantify the individual element
diffusion timescales for sedimentation beneath the convection zones in
our \mesa\ models, and provide tabulated results.
In Section~\ref{S.mixing}, we examine other forms of mixing that can
be relevant in non-convective regions.
These include a greatly refined and expanded treatment of thermohaline
mixing (Section~\ref{S.Thermohaline}), 
as well as convective overshoot (Section~\ref{S.Overshoot}).
Our results confirm the findings of \paperone\ that earth composition
accretion rates approaching
$\dot M_{\rm acc} = 10^{13} \, \rm g \, s^{-1}$ are necessary to match
observed calcium abundances in DA WDs with 
$T_{\rm eff} \gtrsim 15,000 \, \rm K$.
We close with discussion and conclusions in
Sections~\ref{S.discussion}~and~\ref{S.conclusion}.

\section{Surface Convection Zones in Pure~Hydrogen}
\label{S.convection}

Polluting metals quickly mix into the WD surface convection zone,
so that the base of this fully mixed region is where the gravitational
sedimentation rate must be evaluated \citep{Vauclair79,Koester09}.
Here we quantify the total mass, $M_{\rm cvz}$, in the surface
convective layers of our \mesa\ WD models with pure hydrogen
atmospheres and compare to previous work.

To facilitate comparison to the work of \cite{Koester09,Koester10}, we adopt
the ML2 convection prescription \citep{Bohm71} with $\alpha_{\rm MLT}=0.8$.
This value of $\alpha_{\rm MLT}$ is similar to those calibrated against 3D convection
by \cite{Tremblay13,Tremblay15}, but it should be noted
that the calibrated values show some variation with $T_{\rm eff}$.
The depth of the surface
convection zone is also sensitive to the surface boundary condition
of the model. We find that the gray iterative atmosphere procedure in
\mesa\ \citep{Paxton13} provides values of $M_{\rm cvz}$ in good
agreement with \cite{Koester09} for $T_{\rm eff} \gtrsim 9,000\,\rm K$.
At lower effective temperatures, we switch to the pre-computed DA WD
atmosphere tables adapted from \cite{Rohrmann12}. When using these
tables, \mesa\ models attach to boundary conditions given at the optical
depth $\tau_{\rm Ross} = 25$, so the tables are not suitable for
WDs with very shallow convection zones that do not extend to
$\tau_{\rm Ross} > 25$. Fortunately, the gray iterative atmosphere
procedure is adequate for all cases of shallow convection zones, and
it is only necessary to switch to the tables for cooler WDs with large
convection zones. 
For the remainder of this work, we use models that switch from gray
atmosphere boundary conditions to the WD atmosphere tables below
$T_{\rm eff} = 9,000 \, \rm K$.

\begin{figure} 
\begin{center}
\includegraphics[width=\apjcolwidth]{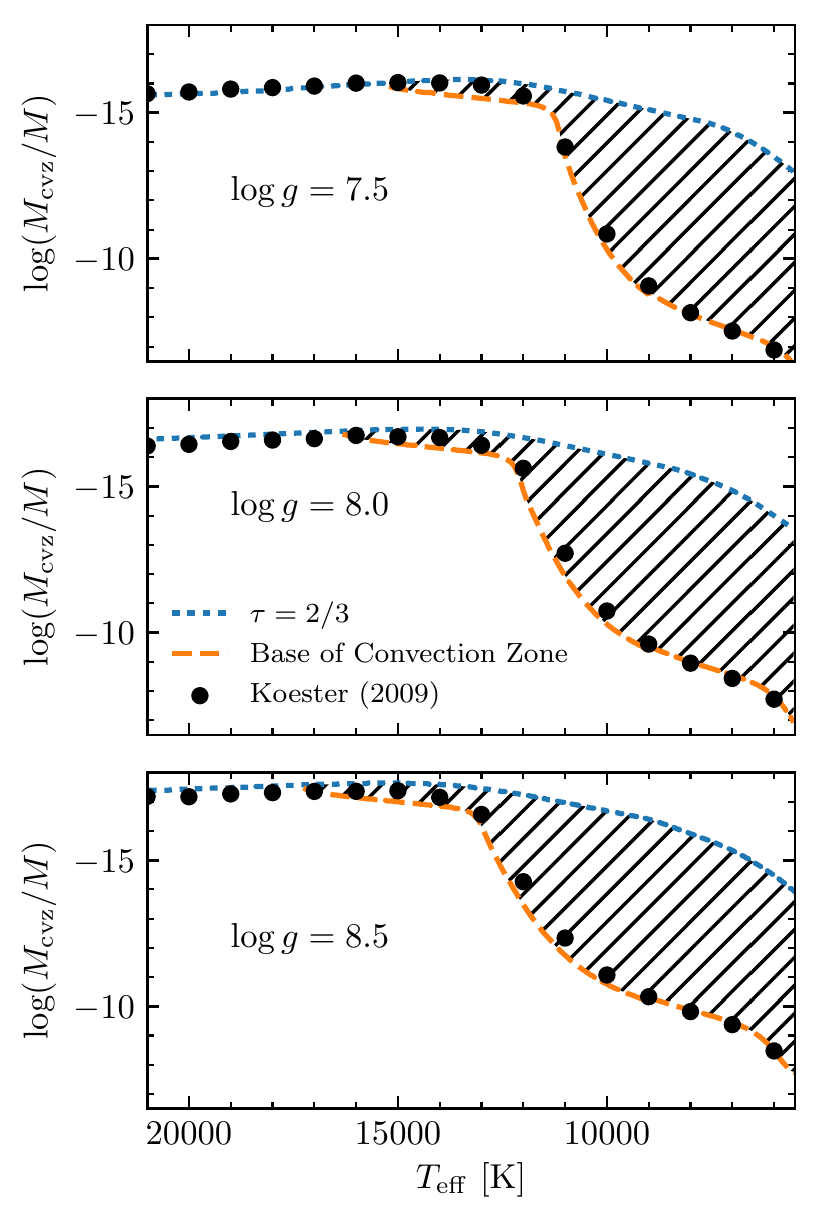}
\caption{
Comparison of the convection zone masses (orange lines) in nearly pure
hydrogen atmospheres of \mesa\ WD models of mass
$0.38 \, M_\odot$, $0.60 \, M_\odot$, and $0.90 \, M_\odot$ to those
given by \citet[black points]{Koester09} for ${\log g = 7.5}$, $8.0$,
and $8.5$.
The blue lines show the location of the photosphere in the \mesa\
models, and the hatched regions indicate the fully mixed convection
zones extending out to the photosphere.
}
\label{fig:cvz}
\end{center}
\end{figure}

Figure~\ref{fig:cvz} shows a comparison of the mass of
the surface convection zone between \mesa\ WD models 
and the DA models of
\cite{Koester09,Koester10}.\footnote{\label{note1} Most recent tables found at
\url{http://www1.astrophysik.uni-kiel.de/~koester/astrophysics/astrophysics.html}.}
For hotter WDs where no surface convection is present ($T_{\rm eff}
\gtrsim 15,000 \, \rm K$), the mass exterior to the photosphere is the
relevant parameter for pollution mixing calculations, so we show this
value as well. The tables of \cite{Koester09} give whichever is
larger of mass in the surface convection zone and mass
exterior to the photosphere.
The hydrogen ionization transition that drives convection results in a
steep change in the convection zone mass around
$10,000 \, {\rm K} \lesssim T_{\rm eff} \lesssim 13,000 \,\rm K$.
We see small disagreements in the exact location of this feature,
and otherwise are in excellent agreement with \cite{Koester09}.
Although the disagreement in mass at fixed
$T_{\rm eff}$ can be up to an order of magnitude, the steep
slope of the curve in this region means that small variations of
$T_{\rm eff}$ within typical observational uncertainties can bring
these values into agreement.

\newpage
\section{Diffusion Timescales}
\label{S.diffusion}

With the structure of \mesa\ WD models and convection zone masses
established, diffusion timescales can now be
calculated for all trace heavy elements. These timescales are
essential to inferring accretion rates and compositions from
observations of trace elements in the photosphere.
Section~2 of \paperone\ shows the equations relating these diffusion
timescales to accretion rates and observable surface abundances. Here
we only repeat a few key definitions for convenience.
When no other mixing occurs beneath the surface convection zone, 
the sedimentation timescale for trace element $i$ is
\begin{equation}
\tau_{{\rm diff},i}\equiv\frac{M_{\rm cvz}}{4\pi r^2\rho v_{{\rm diff},i}}~,
\label{eq:tau_inst}
\end{equation}
where $r$ is the radius, $\rho$ is the density,
and $v_{{\rm diff},i}$ is the sedimentation velocity of element $i$
evaluated at the base of the surface convection zone where
it sinks away from the fully mixed surface region. An approximate
expression for $v_{{\rm diff},i}$ for a trace diffusing element is
given later in Equation~\eqref{eq:trace_vdiff}, but in general our
\mesa\ models calculate diffusion velocities from a full solution of
the Burgers equations \citep{Burgers69} as described in
\cite{Paxton15,Paxton18}.
For a constant accretion rate $\dot M_i$ of species $i$ over
timescales much longer than $\tau_{{\rm diff},i}$, the surface mass
fraction approaches the equilibrium value
\begin{equation}
X_{{\rm eq},i} = \frac{\dot M_i}{M_{\rm cvz}}\tau_{{\rm diff},i}~.
\label{eq:tau_eq}
\end{equation}
For observational inferences, it is assumed that this
equilibrium state has been reached, so that the elemental accretion
rate can be derived from the observed mass fraction as
$\dot M_i = X_{{\rm obs},i} M_{\rm cvz}/\tau_{{\rm diff},i}$. We
denote the total accretion rate as $\dot M_{\rm acc} \equiv \sum_i
\dot M_i$.

\subsection{MESA Diffusion Results}
\label{S.diffusion_results}

The diffusion velocities necessary to calculate diffusion timescales
using Equation~\eqref{eq:tau_inst} are readily available from  \mesa\
models. We obtain these by simply introducing a polluting metal
(e.g.~$^{40}{\rm Ca}$) accreting at a rate of $\dot M_{\rm acc} =
10^{7} \, \rm g \, s^{-1}$. After accretion takes place for many
diffusion timescales, so that the abundance in the surface convection
zone has reached equilibrium, we calculate the diffusion timescale
using Equation~\eqref{eq:tau_inst} along with the diffusion velocity
reported by \mesa\ from the solution of the Burgers equations. These
diffusion calculations include thermal diffusion and properly account
for any degree of electron degeneracy as described in \cite{Paxton18}.

Diffusion calculations according to the Burgers equations rely on
coefficients calculated using a binary scattering formalism.
The well-established coefficients of \cite{Paquette86coeff} are based on a
screened Coulomb potential treatment for calculating binary Coulomb collision
cross sections. Recent updates to \mesa\ \citep{Paxton18}
have included options for using the coefficients
of \cite{Stanton16}, who provide an improvement upon this method with
a more sophisticated treatment of screening.
Table~\ref{tab:Catau} shows some comparisons for diffusion timescales
in a $0.6 \, M_\odot$ WD including calculations using the coefficients
of \cite{Paquette86coeff}. In general, both sets of coefficients give
similar results except for the deepest convection zones, where the
increased Coulomb screening due to electrons in the calculations of
\cite{Stanton16} allows for faster diffusion.

\begin{table*}
\caption{Comparison of \mesa\ and \cite{Koester09} results for the mass of the
  surface convection zone and diffusion timescales for $^{40}{\rm Ca}$
  on a $0.6 \, M_\odot$ WD. The models of \cite{Koester09} adopt a fixed value
  of $\log g = 8.0$, so we include a column for $\log g$ from the 
  \mesa\ model to note how it varies slightly about this
  value as the WD cools.
  Columns labeled with $D_{\rm Paq}$ refer to
  diffusion calculations using the coefficients of
  \cite{Paquette86coeff}, and those labeled $D_{\rm SM}$ refer to
  calculations using the coefficients of \cite{Stanton16}. Columns
  labeled with $Z_2+1$ or $Z_2-1$ refer to calculations for which the
  charge of $^{40}{\rm Ca}$ is taken to be 1 larger or smaller than
  the value given by the ionization calculations based on
  \cite{Paquette86WD}.}
\begin{center}
\begin{tabular}{c | cc | c | ccc | cc | cc}
$T_{\rm eff}$ [K] & \multicolumn{2}{c|}{$\log(M_{\rm cvz}/M)$} & 
$\log g$ &
\multicolumn{7}{c}{$\log (\tau_{\rm diff}/{\rm yr})$ for $^{40}\rm Ca$} \\ 
\hline
 & Koester & \texttt{MESA} & \texttt{MESA} & Koester & \multicolumn{2}{c|}{\texttt{MESA}} & 
\multicolumn{2}{c|}{\texttt{MESA} ($Z_2+1$)} & \multicolumn{2}{c}{\texttt{MESA} ($Z_2-1$)} \\ 
& & & & ($D_{\rm Paq}$) & ($D_{\rm Paq}$) & ($D_{\rm SM}$) & ($D_{\rm Paq}$) & ($D_{\rm SM}$) & ($D_{\rm Paq}$) & ($D_{\rm SM}$) \\
\hline
6000 & -7.722 & -7.8094 & 8.0342 & 4.2924 & 4.2449 & 4.13 & 4.2977 & 4.1827 & 4.1835 & 4.0684 \\
7000 & -8.432 & -8.5222 & 8.0306 & 3.7125 & 3.7107 & 3.6139 & 3.7662 & 3.6695 & 3.6473 & 3.5513 \\
8000 & -8.954 & -8.9849 & 8.0272 & 3.3303 & 3.4113 & 3.3372 & 3.4674 & 3.3923 & 3.3616 & 3.2908 \\
9000 & -9.607 & -9.517 & 8.0238 & 2.8725 & 3.0408 & 2.9957 & 3.1009 & 3.0531 & 2.9715 & 2.9304 \\
10000 & -10.738 & -10.251 & 8.0202 & 1.9997 & 2.476 & 2.4679 & 2.5493 & 2.5371 & 2.3884 & 2.3852 \\
11000 & -12.715 & -11.872 & 8.0164 & 0.4845 & 1.1984 & 1.2236 & 1.3214 & 1.3478 & 1.0337 & 1.0566 \\
12000 & -15.618 & -14.698 & 8.0127 & -1.6941 & -1.0767 & -1.071 & -0.81118 & -0.80264 & -1.5571 & -1.5573 \\
13000 & -16.408 & -16.103 & 8.0094 & -2.359 & -1.9677 & -1.9629 & -1.7151 & -1.7073 & -2.4523 & -2.4534 \\
14000 & -16.672 & -16.292 & 8.006 & -2.6305 & -2.0968 & -2.0931 & -1.8252 & -1.8185 & -2.5713 & -2.5734 \\
15000 & -16.698 & -16.43 & 8.0026 & -2.6277 & -2.1953 & -2.1926 & -1.9216 & -1.9159 & -2.6857 & -2.6887 \\
16000 & -16.744 & -16.622 & 7.9991 & -2.622 & -2.3333 & -2.3318 & -2.0573 & -2.0526 & -2.8272 & -2.8312 \\
17000 & -16.634 & -16.836 & 7.9953 & -2.4688 & -2.4941 & -2.4939 & -2.2153 & -2.2119 & -2.9901 & -2.9952 \\
18000 & -16.586 & -16.787 & 7.9914 & -2.4213 & -2.469 & -2.4694 & -2.1889 & -2.1862 & -2.9668 & -2.9724 \\
19000 & -16.538 & -16.703 & 7.9872 & -2.3804 & -2.4182 & -2.4191 & -2.1374 & -2.1352 & -2.9171 & -2.9231 \\
20000 & -16.439 & -16.644 & 7.983 & -2.3077 & -2.3847 & -2.3862 & -2.1033 & -2.1015 & -2.8849 & -2.8913 \\
\end{tabular}
\end{center}
\label{tab:Catau}
\end{table*}

Table~\ref{tab:Catau} also shows comparisons to the diffusion
timescales given by 
\citet[see link in Footnote~\ref{note1} for the
  most up-to-date diffusion timescale results]{Koester09},
which employ the coefficients of \cite{Paquette86coeff}. When using
these same coefficients, the \mesa\ timescale results agree well as
long as the convection zone depth is comparable. For $T_{\rm eff}
  \gtrsim 11,000 \, \rm K$, the convection zone depths differ by
up to an order of magnitude between \mesa\ and \cite{Koester09},
and the diffusion timescales disagree accordingly.
Table~\ref{tab:SMtau} gives \mesa\ diffusion timescales for ten
commonly observed elements, using the coefficients of
\cite{Stanton16}.

\begin{table*}
\caption{\mesa\ diffusion timescales for the $0.6 \, M_\odot$ WD model
  calculated using the coefficients of \cite{Stanton16}. Supplemental
  tables for other WD masses are available at
  \url{https://doi.org/10.5281/zenodo.2541235} \citep{Zenodo18}.}
\begin{center}
\begin{tabular}{c | c | cccccccccc }
$T_{\rm eff}$ [K] & $\log(M_{\rm cvz}/M)$ &
\multicolumn{10}{c}{$\log (\tau_{\rm diff}/{\rm yr})$} \\
\hline
& & $^{12}{\rm C}$ & $^{16}{\rm O}$ & $^{23}{\rm Na}$ & 
$^{24}{\rm Mg}$ & $^{27}{\rm Al}$ & $^{28}{\rm Si}$ &
$^{40}{\rm Ca}$ & $^{48}{\rm Ti}$ & $^{52}{\rm Cr}$ & 
$^{56}{\rm Fe}$ \\ 
\hline
6000 & -7.8094 & 4.2573 & 4.3303 & 4.0949 & 4.0564 & 4.001 & 3.8702 & 4.13 & 4.0304 & 3.931 & 3.89 \\
7000 & -8.5222 & 3.8338 & 3.9012 & 3.646 & 3.6249 & 3.4721 & 3.4464 & 3.6139 & 3.5067 & 3.4046 & 3.4617 \\
8000 & -8.9849 & 3.5523 & 3.6214 & 3.3551 & 3.3105 & 3.1875 & 3.1688 & 3.3372 & 3.1957 & 3.0915 & 3.185 \\
9000 & -9.517 & 3.2304 & 3.2723 & 2.9774 & 2.935 & 2.8664 & 2.8485 & 2.9957 & 2.8552 & 2.7713 & 2.8289 \\
10000 & -10.251 & 2.7752 & 2.7551 & 2.4842 & 2.4591 & 2.2849 & 2.3975 & 2.4679 & 2.3434 & 2.3276 & 2.3132 \\
10500 & -10.928 & 2.3481 & 2.2194 & 1.9873 & 1.8792 & 1.8257 & 1.973 & 1.9514 & 1.8086 & 1.9119 & 1.8567 \\
11000 & -11.872 & 1.7424 & 1.552 & 1.2795 & 1.0427 & 1.2254 & 1.3768 & 1.2236 & 1.1432 & 1.233 & 1.184 \\
11500 & -13.147 & 0.74402 & 0.57918 & 0.21533 & 0.17977 & 0.38994 & 0.54935 & 0.22032 & 0.31781 & 0.28456 & 0.21358 \\
12000 & -14.698 & -0.55196 & -0.70814 & -1.3172 & -0.85012 & -0.63472 & -0.65315 & -1.071 & -0.88082 & -0.9153 & -0.94716 \\
12500 & -15.953 & -1.6971 & -1.9663 & -2.124 & -1.6439 & -1.6901 & -1.7108 & -1.8661 & -1.9454 & -1.9789 & -2.0115 \\
13000 & -16.103 & -1.9121 & -2.0535 & -2.2121 & -1.74 & -1.8271 & -1.8073 & -1.9629 & -2.0423 & -2.0772 & -2.1094 \\
13500 & -16.203 & -1.9954 & -2.1254 & -2.2838 & -1.809 & -1.9515 & -1.8764 & -2.0319 & -2.1114 & -2.1462 & -2.1785 \\
14000 & -16.292 & -2.0585 & -2.1858 & -2.3321 & -1.87 & -2.0932 & -1.9374 & -2.0931 & -2.1725 & -2.2074 & -2.2272 \\
15000 & -16.43 & -2.1621 & -2.2885 & -2.4473 & -1.9693 & -2.2683 & -2.0368 & -2.1926 & -2.2721 & -2.3071 & -2.3393 \\
16000 & -16.622 & -2.3042 & -2.4307 & -2.5896 & -2.1081 & -2.6023 & -2.1757 & -2.3318 & -2.4114 & -2.4464 & -2.4786 \\
17000 & -16.836 & -2.4658 & -2.5932 & -2.7529 & -2.2694 & -2.8193 & -2.3373 & -2.4939 & -2.5737 & -2.6088 & -2.6411 \\
18000 & -16.787 & -2.4429 & -2.5704 & -2.7301 & -2.2449 & -2.6484 & -2.3128 & -2.4694 & -2.5493 & -2.5843 & -2.6167 \\
19000 & -16.703 & -2.3906 & -2.5212 & -2.6809 & -2.1946 & -2.2487 & -2.2625 & -2.4191 & -2.4989 & -2.5339 & -2.5663 \\
20000 & -16.644 & -2.261 & -2.4894 & -2.6491 & -2.1617 & -2.2136 & -2.2296 & -2.3862 & -2.466 & -2.5009 & -2.5333
\end{tabular}
\end{center}
\label{tab:SMtau}
\end{table*}

\subsection{Approaching Equilibrium}

\begin{figure}
\begin{center}
\includegraphics[width=\apjcolwidth]{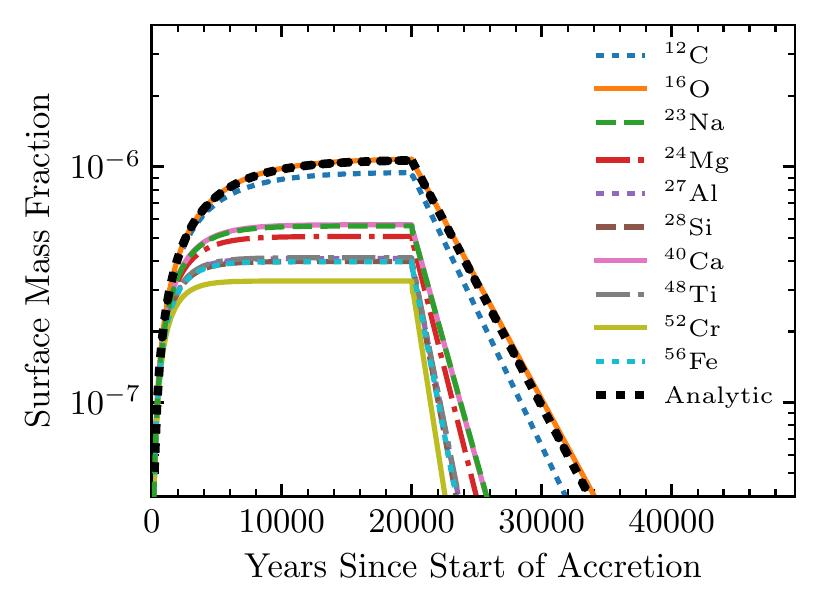}
\caption{
  Mass fractions over time in a $0.6 \, M_\odot$ WD model at ${T_{\rm eff}
  = 8,000 \, \rm K}$ that accretes 10 isotopes at a rate of $10^7 \,
  \rm g \, s^{-1}$ each for $20,000$ years, after which accretion
  ends and the pollutants sediment away from the surface. The
  black dashed curve shows the analytic solution for $^{16} \rm O$
  with $\log(\tau_{\rm diff}/\rm yr) = 3.62$.
}
\label{fig:turnoff}
\end{center}
\end{figure}
\begin{figure} [b]
\begin{center}
\includegraphics[width=\apjcolwidth]{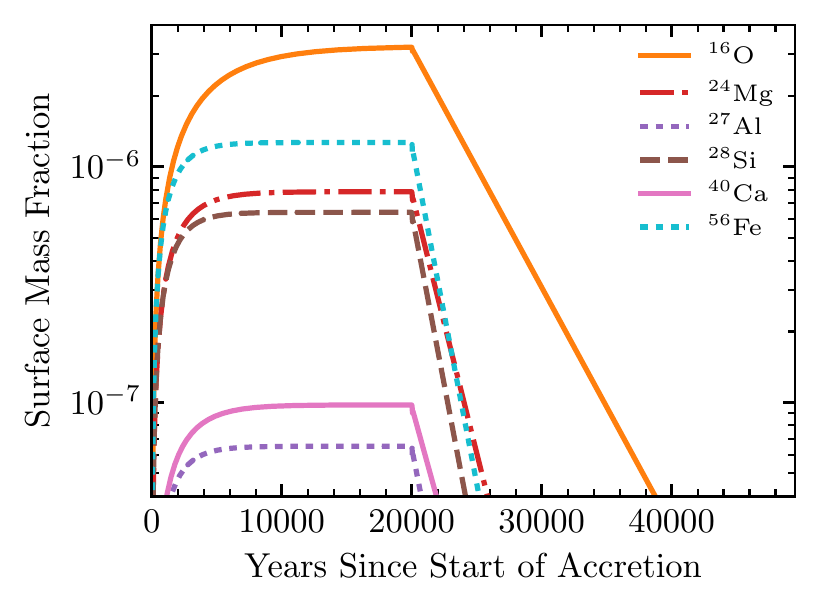}
\caption{
  Mass fractions over time in a $0.6 \, M_\odot$ WD model at ${T_{\rm eff}
  = 8,000 \, \rm K}$ that accretes at a total rate of
  $10^{8} \, \rm g \, s^{-1}$ with bulk earth composition for
  $20,000$ years, after which accretion ends and the pollutants
  sediment away from the surface. Only the most abundant elements
  appear on the scale shown here.
}
\label{fig:turnoff_earth}
\end{center}
\end{figure}
Figure~\ref{fig:turnoff} shows surface mass fractions for several accreting
elements in a \mesa\ model, first approaching equilibrium after
accretion turns on and continues for many diffusion timescales, then
sinking away after accretion shuts off. For comparison, this figure
also shows the analytic solution described in \paperone\  for this
constant accretion rate for $^{16} \rm O$ with a diffusion
timescale of $\log(\tau_{\rm diff}/\rm yr) = 3.62$. This verifies that
the metals approach the equilibrium surface mass fraction
predicted by Equation~\eqref{eq:tau_eq} for the diffusion timescales
given in Table~\ref{tab:SMtau}.
The accretion episode shown in Figure~\ref{fig:turnoff} has all
elements accreting at equal rates ($10^7 \, \rm g \, s^{-1}$ for each
element) to illustrate the effects of the hierarchy of diffusion timescales.
This manifests as a clear ordering of abundances, where
those with the longest diffusion timescales appear as the most
abundant over all phases. In contrast,
Figure~\ref{fig:turnoff_earth} shows a more realistic accretion
scenario, where the elements accrete at the total rate
$\dot M_{\rm acc} = 10^8 \, \rm g \, s^{-1}$, but with the bulk
earth abundance ratios of \cite{McD}.
In this case, both the relative accreted mass
fraction and diffusion timescale for each element play a role in
establishing the final hierarchy of observed surface abundances.
Neither of these calculations include thermohaline mixing (see
Section~\ref{S.Thermohaline}).

\subsection{Ionization States for Trace Metals}

\begin{figure*} 
\begin{center}
\includegraphics[width=\apjcolwidth]{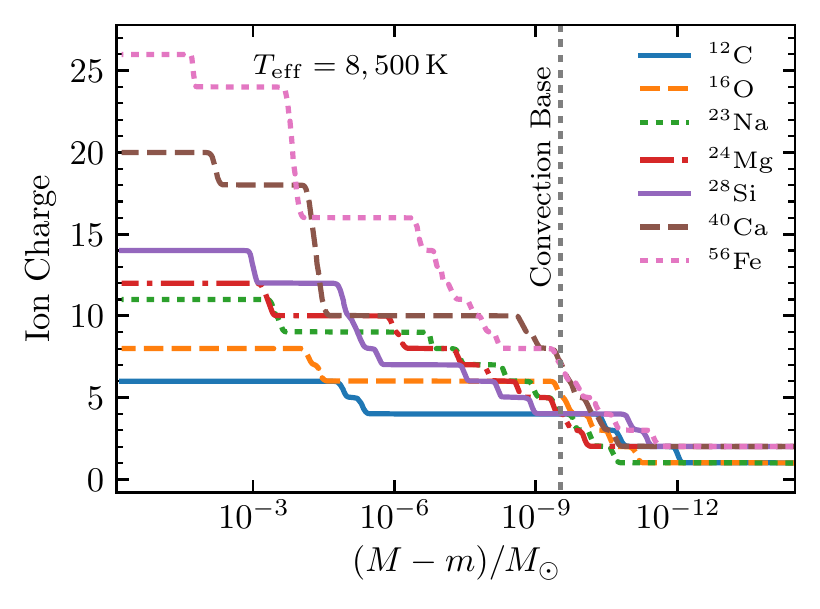}
\hspace{2em}
\includegraphics[width=\apjcolwidth]{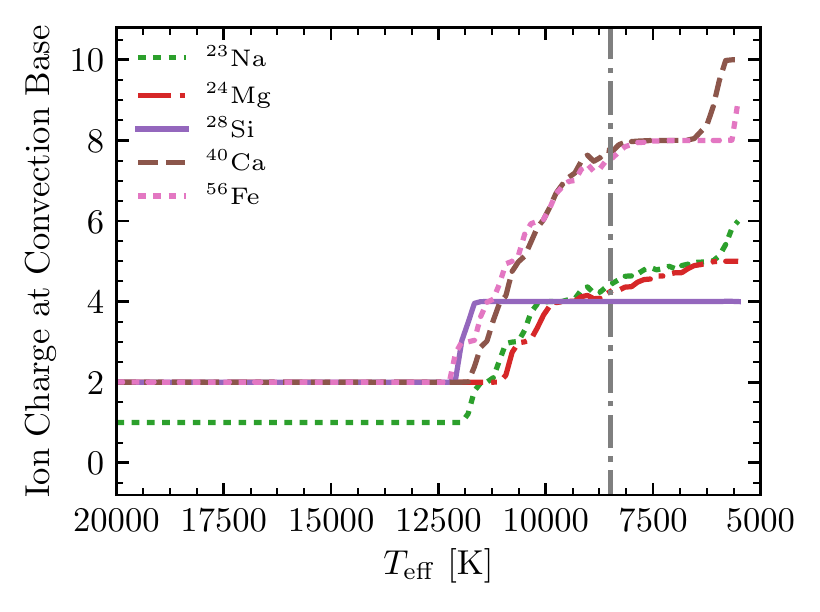}
\caption{
  {\it Left:} Profiles of average ion charges in the outer layers of a
  $0.6 \, M_\odot$ WD at $T_{\rm eff} = 8,500 \, \rm K$.
  {\it Right:} Ion charges at the base of the convection zone as a function of WD
  temperature for a $0.6 \, M_\odot$ WD. The gray dashed line indicates
  the temperature of the model shown in the left panel.
}
\label{fig:ion_profile}
\end{center}
\end{figure*}

The partial ionization of metals in the surface layers relevant for
pollution has important effects for the diffusion timescales.
If we denote the background material in which diffusion takes place by the
index~$1$ (hydrogen in the case of a DA WD atmosphere), and denote the
pollutant by index~$2$, then in the limit of a trace pollutant ($n_2
\ll n_1$) its diffusion velocity can be expressed as
(cf.~\citealt{Pelletier86,Dupuis92,Koester14})
\begin{equation}
\begin{aligned}
v_{\rm diff} = D_{12} 
\bigg[ &- \frac{\partial \ln c_2}{\partial r} + \left(
  \frac{Z_2}{Z_1}A_1 - A_2 \right) \frac{m_{\rm p}g}{kT} \\
& + \left( \frac{Z_2}{Z_1} -1 \right) \frac{\partial \ln p_i}{\partial r} 
+ \alpha_T \frac{\partial \ln T}{\partial r}\bigg]~,
\end{aligned}
\label{eq:trace_vdiff}
\end{equation}
where $c_2 \equiv n_2/(n_1+n_2)$ is the concentration of the
pollutant, $p_i \equiv p_1 + p_2$ is the ion pressure, $\alpha_T$ is
the thermal diffusion coefficient, and $A$ and $Z$ refer to the mass
and charge of each species respectively.
Note that this equation is appropriate for any degree of electron
degeneracy \citep{Pelletier86}, and it agrees with our \mesa\
diffusion treatment based on the Burgers equations in the limit of
trace particles diffusing in a hydrogen background.
The charge of each species influences the diffusion velocity in two
important ways: the direct influence on the forcing terms felt by each
ion seen in Equation~\eqref{eq:trace_vdiff}, and the influence of the
charge of each particle on the Coulomb scattering that results in the
diffusion coefficient $D_{12}$.
The diffusion coefficient is related to the resistance coefficients
used for \mesa\ diffusion calculations described in
\cite{Paxton15,Paxton18} by $D_{ij} \propto K_{ij}^{-1}$. For Coulomb
collisions, the resistance coefficients described in \cite{Paxton15}
scale with the charge approximately as $K_{ij} \propto (Z_i Z_j)^2$, and
hence diffusion calculations can be very sensitive to the ionization
treatment adopted for the partially ionized surface regions of WD
models.

Formally, each ionization state of a given element may be treated as a
separate species with its own integer charge $Z_i$ for purposes of
diffusion calculations. In order to simplify the problem, \mesa\
calculations instead adopt an average state $\bar Z$ for each element
as described in \cite{Paxton15,Paxton18} so that each isotope
corresponds to only one diffusion species. 
We use the ionization treatment of \cite{Paquette86WD} to find an
average charge state for each diffusion species everywhere in the
\mesa\ model.\footnote{We note that the expression in \cite{Paquette86WD}
for the depression of the continuum for ionization potentials contains
a typo in Equation~(21), where a factor of $\rho^{1/3}$ is missing
from the last line.
The \mesa\ ionization routine instead follows Equation~(3) of
\cite{Dupuis92}, which correctly includes this factor.
Our ionization treatment is very similar to that of \cite{Koester09},
who also notes correcting the missing factor of $\rho^{1/3}$ for the
most recent calculations hosted on his website (see link
  in Footnote~\ref{note1}).}

Figure~\ref{fig:ion_profile} displays some of the
charges used as input for diffusion calculations reported in
\mesa\ WD models. Since the ionization procedure based on \cite{Paquette86WD}
involves comparing ionization potentials to an effective threshold  potential, it
always selects an integer value for the average charge. This results
in the stair-stepped profiles seen in Figure~\ref{fig:ion_profile},
which have been smoothed slightly to improve the numerical stability of
diffusion calculations. The last columns of Table~\ref{tab:Catau}
present results from diffusion calculations for which the charge $Z_2$
is taken to be one larger or smaller than the value obtained from the
\cite{Paquette86WD} routine. Comparison of these timescales
quantifies the rough uncertainty associated with the average
ionization calculations here. 

Our diffusion calculations assume that every species is at least singly
ionized. Diffusion coefficients for a neutral species require
collision integrals for dipole scattering, which result in
significantly smaller collision cross sections and correspondingly
faster diffusion timescales (Appendix~\ref{S.neutral}).
Options for such diffusion coefficients
are not currently available in \mesa. Since diffusion fluxes for
neutral elements can be much faster than those for singly ionized
elements, even a small fraction of neutral particles in the relevant
layer can significantly modify overall sedimentation timescales, and
it is no longer appropriate to treat ionization with an average charge
$\bar Z < 1$. Diffusion timescales presented in this work are only
accurate for models where surface temperatures are
hot enough or surface convection zones reach depths sufficient for at
least single ionization of pollutants. Due to a thin or absent surface
convection zone, these conditions fail to be
satisfied around  $T_{\rm eff} \approx 15,000 \, \rm K$,
and corresponding disagreement is evident between our results
and those of \cite{Koester09} in Table~\ref{tab:Catau}
for this regime.

\section{Other Mixing}
\label{S.mixing}

We now explore additional mixing other than element diffusion beneath
the convective layer. We focus on two fluid processes that
can cause additional mixing: the thermohaline instability and convective
overshoot.

\subsection{Thermohaline Mixing}
\label{S.Thermohaline}

In the context of WD pollution, \cite{Deal13} were the first to
explore the possibility that accreted metals in WD atmospheres may
lead to thermohaline instability. Subsequent work by \cite{Wachlin17}
confirmed the importance of the resulting mixing.
In \paperone, we extended the parameter
space for polluted WDs where thermohaline instability may occur,
finding that thermohaline mixing significantly
modifies inferred accretion rates in hydrogen-atmosphere WDs
with $T_{\rm eff} \gtrsim 10,000 \, \rm K$, with some rates reaching
${\dot M_{\rm acc} \approx 10^{13} \, \rm g \, s^{-1}}$ for
$T_{\rm eff} \gtrsim 15,000 \, \rm K$. Our exploration in
\paperone\ was limited to WD models of mass $M = 0.6 \, M_\odot$
($\log g = 8.0$). We now expand upon that work with models of other
masses to allow interpolation in $\log g$. We also adopt a refined
treatment of thermohaline mixing based upon the work of
\cite{Brown13}, which is calibrated against 3D simulations.

Two criteria must be satisfied for the thermohaline instability to be
active. First, there must be an inverted molecular weight gradient in
a region that is stable to convection:
\begin{equation}
\nabla_T - \nabla_{\rm ad} < \frac{\varphi}{\delta} \nabla_\mu < 0~,
\label{eq:ineq1}
\end{equation}
where $\nabla_T = (\partial \ln T/\partial \ln P)$
is the temperature gradient in the fluid,
$\nabla_{\rm ad} = (\partial \ln T / \partial \ln P)_s$
is the adiabatic temperature gradient, 
${\nabla_\mu = (\partial \ln \mu/\partial \ln P)}$
is the mean molecular weight gradient,
${\varphi = (\partial \ln \rho/\partial \ln \mu)_{P,T}}$, 
and ${\delta = - (\partial \ln \rho/ \partial \ln T)_{P,\mu}}$.
The instability is then driven by thermal exchange of perturbed
fluid elements with their surroundings, whereupon a density
contrast due to $\nabla_\mu$ leads to further mixing.
Thus, the second criterion for thermohaline instability is that the
magnitude of the molecular weight gradient $\nabla_\mu$ and the
thermal diffusivity $\kappa_T$ must be large enough to excite the
instability before particle diffusivity $\kappa_\mu$ within a perturbed fluid element
can adjust its composition \citep{Baines69,Garaud18}:
\begin{equation}
\frac{(\varphi/\delta) \nabla_\mu}{\nabla_T - \nabla_{\rm ad}} > \frac{\kappa_\mu}{\kappa_T}~.
\label{eq:ineq2}
\end{equation}
Assuming that heat transport is radiative, the thermal diffusivity is
\begin{equation}
\kappa_T = \frac{4 a c T^3}{3 \kappa \rho^2 C_P}~,
\end{equation}
where $\kappa$ is the opacity and $C_P$ is the heat capacity.
The particle diffusivity is derived from the diffusion
coefficients of the various polluting metals that determine the
molecular weight of a fluid element. 

The mixing that results from thermohaline instability can be
approximated with a coefficient that scales with thermal diffusivity
and the molecular weight gradient \citep{KRT80}:
\begin{equation}
D_{\rm th} = \alpha_{\rm th} \kappa_T \frac{3}{2}
\frac{(\varphi/\delta) \nabla_\mu}{\nabla_T - \nabla_{\rm ad}}~,
\label{eq:KRTD}
\end{equation}
where $\alpha_{\rm th}$ is a dimensionless efficiency parameter. In
\paperone, we explored inferences for polluted WD accretion rates
using the mixing treatment of Equation~\eqref{eq:KRTD} with
$\alpha_{\rm th} = 1$. Note that this
mixing treatment does not explicitly check the criterion for
instability given in Equation~\eqref{eq:ineq2}, but in \paperone\ 
we verified that it is satisfied for regions of interest
in polluted WDs where thermohaline mixing may occur.

\begin{figure} 
\begin{center}
\includegraphics[width=\apjcolwidth]{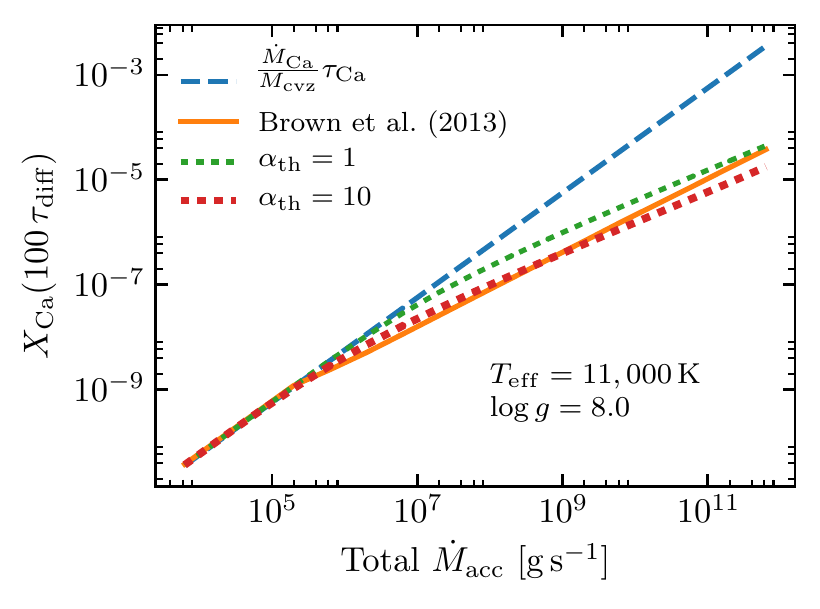}
\caption{
Surface Ca mass fraction after 100 diffusion timescales as a function
of total metal accretion rate for a $0.6 \, M_\odot$ ($\log g = 8.0$)
WD model accreting metals in bulk earth ratios. Curves labeled with
values of $\alpha_{\rm th}$ use the thermohaline mixing prescription
of Equation~\eqref{eq:KRTD}, while the orange curve employs
the \cite{Brown13} routine in \mesa\ version 11191. The blue dashed
line shows the expectation according to Equation~\eqref{eq:tau_eq} if
the diffusive sedimentation timescale governs surface abundances.
}
\label{fig:thermo_eff}
\end{center}
\end{figure}

\mesa\ also offers a thermohaline mixing treatment based on the work
of \cite{Brown13}, which is calibrated against their 3D hydrodynamic
simulations.
This treatment explicitly accounts for the criterion in
Equation~\eqref{eq:ineq2} and produces a mixing coefficient
designed to scale smoothly to zero as conditions approach the limit
defined there.
Figure~\ref{fig:thermo_eff} shows the surface Ca mass fraction in
polluted $0.6 \, M_\odot$ WD models after accreting for many
diffusion timescales, with thermohaline mixing according to either
Equation~\eqref{eq:KRTD} or \cite{Brown13}.
These \mesa\ models also include element diffusion at all times. 
Figure~\ref{fig:thermo_eff} shows that results for polluted WD models 
using thermohaline mixing based on \cite{Brown13} are qualitatively
similar to those using Equation~\eqref{eq:KRTD} with
$1 \lesssim \alpha_{\rm th} \lesssim 10$. 

\begin{table}
\caption{Critical accretion rates in our $0.6 \, M_\odot$ \mesa\ WD
  models above which thermohaline mixing modifies the equilibrium
  surface mass fractions from the prediction of
  Equation~\eqref{eq:tau_eq}. This assumes accretion of material with
  bulk earth composition \citep{McD}.
} 
\begin{center}
\begin{tabular}{c | cccc}
$T_{\rm eff}$ [K] & 6,000 & 7,000 & 8,000 & 9,000 \\
\hline
$\dot M_{\rm crit}$ [$\rm g \, s^{-1}$] & $10^{12}$ & $10^{10}$ & $10^{9}$ & $10^{8}$  
\\ 
\hline
\hline
\hline
$T_{\rm eff}$ [K] & 10,000 & 11,000 & \multicolumn{2}{c}{$>12$,000} \\
\hline
$\dot M_{\rm crit}$ [$\rm g \, s^{-1}$] & $10^{7}$ & $10^{6}$ & \multicolumn{2}{c}{$<10^{4}$} 
\end{tabular}
\end{center}
\label{tab:critical}
\end{table}

For small accretion rates, thermohaline mixing is not
active, and the equilibrium surface mass fractions shown in
Figure~\ref{fig:thermo_eff} match the
prediction of Equation~\eqref{eq:tau_eq} for the diffusion timescales
given in Table~\ref{tab:SMtau}. Above a critical accretion rate $\dot
M_{\rm crit}$, the metal concentration at the surface builds up a
sufficient magnitude of $\nabla_\mu$ to excite thermohaline
instability, and \mesa\ models including thermohaline mixing diverge
from the prediction of Equation~\eqref{eq:tau_eq}.
For a ${T_{\rm eff} = 11,000 \, \rm K}$ WD,
Figure~\ref{fig:thermo_eff} shows that this critical rate is around
$\dot M_{\rm crit} \approx 10^6 \, \rm g \, s^{-1}$.
Table~\ref{tab:critical} gives values of this critical rate for \mesa\
WD models over a range of $T_{\rm eff}$. 
For models with ${T_{\rm eff} > 12,000 \, \rm K}$, the surface
convection zones are so small that thermohaline
mixing is active for all accretion rates in the range that we
explored ($\dot M_{\rm acc} > 10^4 \, \rm g \, s^{-1}$).

The curve shown for the \cite{Brown13} prescription in
Figure~\ref{fig:thermo_eff} varies slightly from the similar plot
shown in Figure~3 of \paperone. This is due to a small correction to
the \mesa\ implementation of this routine that affects the mixing
coefficient in the regime near the limit of thermohaline
instability. This correction was introduced after \mesa\ release
version 10398, which was used for \paperone, but it is present in \mesa\
version 11191, which we use for all models that include thermohaline
mixing in this paper.
The asymptotic analysis regimes presented in
Appendix~B of \cite{Brown13} form the basis of the 1D mixing
treatment. In particular, their Appendix~B.3 addresses the regime in
which the fluid is near the limit imposed by Equation~\eqref{eq:ineq2}. 
The method relies on an expansion in the parameter
\begin{equation}
\epsilon \equiv 1 - 
\frac{\kappa_\mu/\kappa_T}{(\varphi/\delta)\nabla_\mu/(\nabla_T -
  \nabla_{\rm ad})}~,
\end{equation}
which is assumed to be small. The implementation for this regime in
\mesa\ version 11191 ensures that this parameter is
sufficiently small whenever applying the method of \cite{Brown13}
Appendix~B.3, yielding more consistent results than version 10398.
With these corrections, the \mesa\ implementation shows more
mixing near the boundary of thermohaline instability defined by
Equation~\eqref{eq:ineq2}. Hence, models
employing the \cite{Brown13} routine in \mesa\ version 11191
diverge from the prediction of diffusion alone at the lower accretion
rates seen in Figure~\ref{fig:thermo_eff}.

\begin{figure} 
\begin{center}
\includegraphics[width=\apjcolwidth]{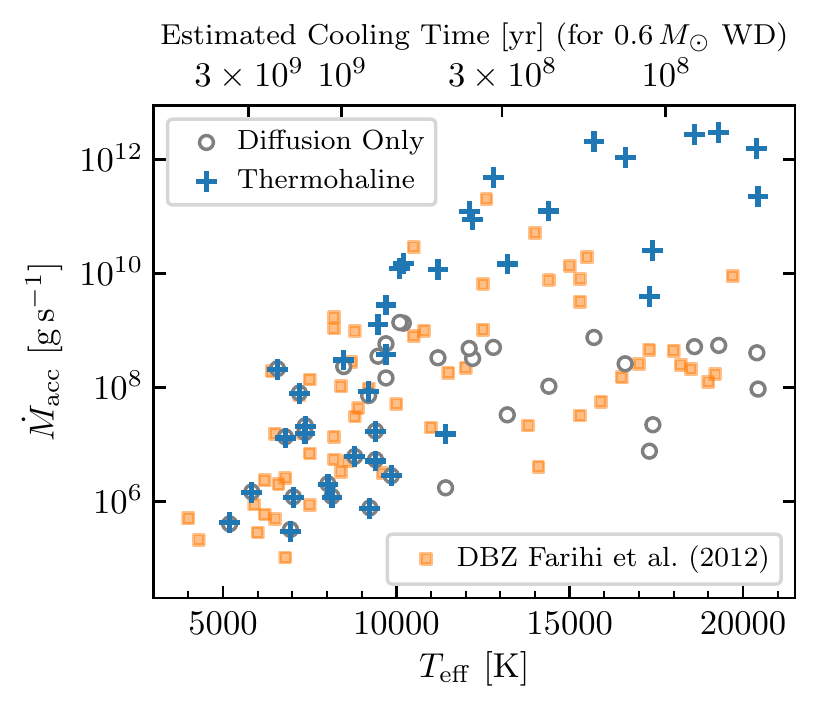}
\caption{
Accretion rates inferred with (blue crosses) and without (open circles)
thermohaline mixing from the observed Ca abundances for the 38 WDs
given in Table~1 of \cite{Koester06}. Models that include thermohaline
mixing follow the prescription of \cite{Brown13}.
Accreting material is assumed to have bulk earth composition.
The orange points show the rates inferred for He-atmosphere WDs by
\cite{Farihi12} for comparison.
The top axis shows estimated WD cooling time based on a \mesa\ model
of a $0.6 \, M_\odot$ DA WD.
}
\label{fig:thermo_mdot}
\end{center}
\end{figure}

Using this updated thermohaline mixing treatment, we construct a large
grid of accreting DA WD models as in \paperone.
Effective temperatures of the models span the range
${6,000 \, {\rm K} < T_{\rm eff} < 20,500 \,\rm K}$,
and accretion rates for each temperature span
$10^4 \, {\rm g \, s^{-1}} < \dot M_{\rm acc} < 10^{12} \, \rm g \, s^{-1}$. 
All models accrete bulk earth material \citep{McD}.
We tabulate values of $X_{\rm Ca}$ present at the surface of each model
after 100 diffusion timescales as defined by Table~\ref{tab:SMtau}. We
then interpolate on these tables to map observed values of
$X_{\rm Ca}$ to total inferred accretion rates $\dot M_{\rm acc}$. We
also expand upon the results of \paperone\ by providing these tables
for models with three different WD masses to allow interpolation
in $\log g$: $M_{\rm WD}/M_\odot = 0.38, 0.60, 0.90$ 
(${\log g \approx 7.5, 8.0, 8.5}$). These tables
are available along with simple python interpolation routines
at \url{https://doi.org/10.5281/zenodo.2541235} \citep{Zenodo18}.

Figure~\ref{fig:thermo_mdot} shows inferred accretion rates based on
these tables for the same sample of polluted DA WDs \citep{Koester06} that was
discussed in \paperone. In general, accretion rates are similar to the
inferences made in \paperone, though the very highest inferences are
slightly lower than the previous highest values.
A few WDs also show adjustments due to observed
values of $\log g$ different from the value of $8.0$ assumed in
\paperone, especially for
${9,000 {\, \rm K} < T_{\rm eff} < 13,000 \, \rm K}$, 
where the surface convection zone masses are especially
sensitive to $\log g$ (see Figure~\ref{fig:cvz}).
However, the overall qualitative picture remains the same.
Thermohaline mixing causes inferred accretion rates to increase by 
several orders of magnitude for WDs with
${T_{\rm eff} > 10,000 \, \rm K}$!

\subsubsection{Non-constant Accretion Rates}

The previous section shows results when accretion
occurs in a steady state for many diffusion timescales, allowing the
surface metal pollution to approach equilibrium abundances. However, if
the source of accretion supplied to the surface varies with time, this
can introduce complexities in the $\nabla_\mu$ profile that sets the
conditions for thermohaline instability according to
Equations~\eqref{eq:ineq1}~and~\eqref{eq:ineq2}.
In particular, heavy elements must be continually supplied to the
surface to maintain $\nabla_\mu < 0$ in the mixing region relevant to
observable pollution.
If the accretion rate decreases significantly, the gradient necessary
for thermohaline instability can disappear, halting thermohaline
mixing.

\begin{figure} 
\begin{center}
\includegraphics[width=\apjcolwidth]{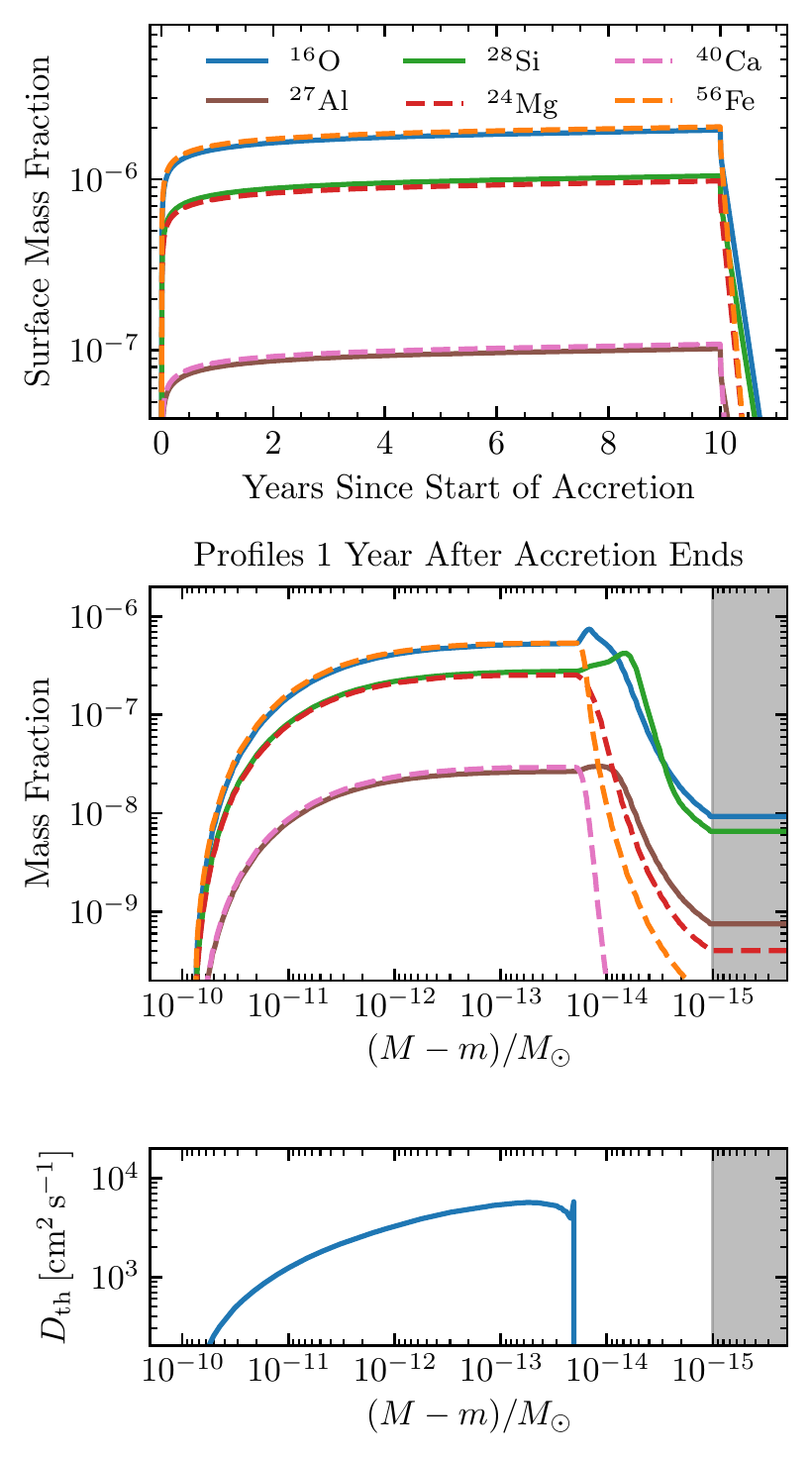}
\caption{
A $0.6 \, M_\odot$, $T_{\rm eff} = 12,000 \, \rm K$ \mesa\ DA WD model
including thermohaline mixing. This model accretes at a rate of
$10^8 \, \rm g \, s^{-1}$ for
$10 \, {\rm yr} \approx 100 \tau_{\rm diff}$,
after which accretion ends and metals sediment away from the photosphere.
The lower panels show the interior profile of the model one year after
accretion has ceased.
The gray shaded region represents the fully mixed surface convection
zone.
}
\label{fig:NonConst}
\end{center}
\end{figure}

As a simple illustration, we show in Figure~\ref{fig:NonConst} a \mesa\ model
including thermohaline mixing that accretes at a constant rate until
it approaches equilibrium, followed by a cessation of accretion after
$10 \, {\rm yr} \approx 100 \tau_{\rm diff}$.
Due to the sudden disappearance of an inverted
$\nabla_\mu$ in the surface region governing observable metal
pollution, thermohaline mixing is no longer relevant. Instead, the
diffusion timescales of Table~\ref{tab:SMtau} dictate the fast
exponential decay of metal pollution at the photosphere.
These results contrast with the diffusion-only \mesa\ model shown in
Figure~\ref{fig:turnoff_earth}, where the same diffusion timescale
governs both the approach to equilibrium and exponential decay after
accretion ceases.

Figure~\ref{fig:NonConst} also demonstrates important features involving
differentiation of the accreted composition.
When thermohaline mixing is active near the
surface during the constant accretion phase, no composition differentiation
occurs because fluid elements that dominate the mixing transport all
elements together.
However, once thermohaline mixing ceases near the surface, individual
particle diffusion dominates, and significant differentiation
quickly occurs within a few diffusion timescales.
The middle panel of Figure~\ref{fig:NonConst} shows
that the deeper layers where thermohaline mixing is still active
reflect the accreted bulk earth composition \citep{McD}, but separate
diffusion timescales for each element quickly rearrange the surface
composition.
Elements with the shortest diffusion timescales such as
$^{56}{\rm Fe}$ disappear from the surface much sooner, even when they
were previously among the most abundant due to the accreted composition.

\subsubsection{Helium-dominated Atmospheres}

WDs with helium-dominated atmospheres do not experience the same
corrections due to thermohaline mixing that hydrogen-dominated
atmospheres do.
Two effects conspire to greatly reduce the potential for a
$\nabla_\mu$ large enough to excite thermohaline instability. First,
the mean molecular weight of the dominant background material (He) is
more than double that in the case of a hydrogen atmosphere, so the
contrast with accreting metals is not as severe. Second, surface
convection zones for helium atmospheres contain much more mass than
hydrogen at a given temperature \citep{Koester09}. This dilutes
accreted metals and prevents the buildup of a significant $\nabla_\mu$
below the convection zone.

\begin{figure} 
\begin{center}
\includegraphics[width=\apjcolwidth]{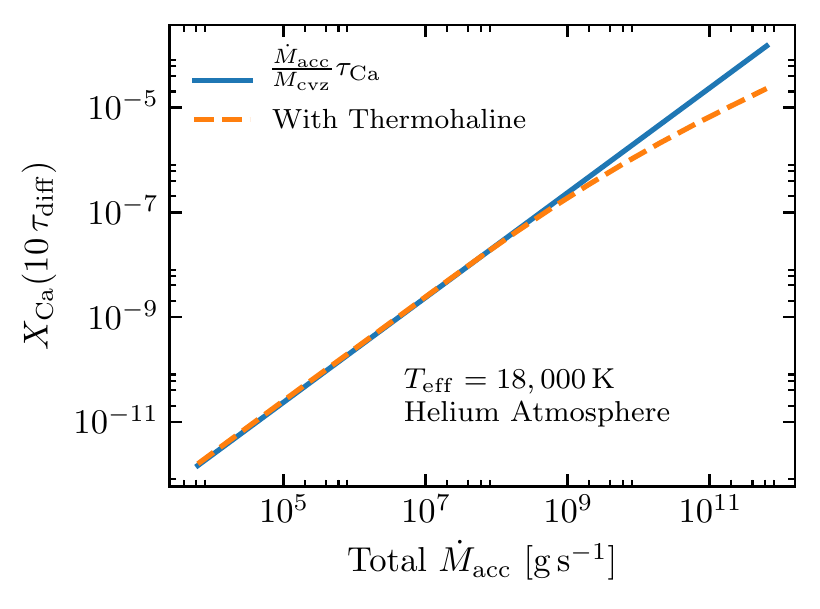}
\caption{
Surface Ca mass fraction after accreting bulk earth material for
10 diffusion timescales for a $T_{\rm eff} = 18,000 \, \rm K$ \mesa\
WD model with a helium-dominated atmosphere.
}
\label{fig:DB}
\end{center}
\end{figure}

For example, we constructed a $0.59 \, M_\odot$ \mesa\ WD model with
$T_{\rm eff} = 18,000 \, \rm K$ and a pure He atmosphere. We found
that the surface convection zone mass of this model was $M_{\rm cvz}
= 8 \times 10^{-8} \, M_\odot$, and diffusion timescales for accreted
metals were on the order of $10^5$ years. These values
agree with the tables of \cite{Koester09} for $\log g = 8.0$ DB WDs.
We explored \mesa\ runs for this WD
model accreting bulk earth composition at rates in the range
$10^4 \, {\rm g \, s^{-1}} < \dot M_{\rm acc} < 10^{12} \, \rm g \,
s^{-1}$. We included thermohaline mixing in the runs using the
treatment of Equation~\eqref{eq:KRTD} with $\alpha_{\rm th} = 10$ (the
\mesa\ treatment based on \cite{Brown13} is not applicable here
because it assumes a hydrogen-dominated
background). Even for the highest accretion rates, we find adjustments
of at most one order of magnitude to inferred accretion rates compared to
calculations that assume no thermohaline mixing (Figure~\ref{fig:DB}). 
Figure~\ref{fig:thermo_mdot} shows that typical accretion rates
inferred for DB WDs in this temperature range are $10^8$--$10^{10} \,
\rm g \, s^{-1}$, and our \mesa\ models show negligible corrections
due to thermohaline mixing in this regime. 
The surface convection zone grows up to three orders of magnitude 
larger for cooler WDs \citep{Koester09}, and the largest rates
inferred for DB WDs only reach $10^{11} \, \rm g \, s^{-1}$, 
so thermohaline mixing will be inconsequential for
He-atmosphere WDs with $T_{\rm eff} \lesssim 18,000 \, \rm K$.

Below $T_{\rm eff} \approx 16,000 \, \rm K$, the cool, dense, neutral
helium at the surface of the WD falls outside the regime covered by opacity
tables currently available in \mesa\ \citep{Paxton11}.
The code is therefore not able to set a physical outer boundary
condition for the model below this temperature. Tabulated outer
boundary conditions such as those used in the case of
hydrogen-dominated atmospheres (Section~\ref{S.convection}) have been
implemented in other WD codes \citep[e.g.,][]{Camisassa17}, but no
such option is currently available in \mesa. In the context of
polluted WDs, it is also unclear whether atmosphere conditions
tabulated for pure helium would be sufficient,
since opacity may be sensitive to contaminating metals through effects
such as He$^-$ free-free absorption. Without the ability to set an
appropriate outer boundary condition, \mesa\ models cannot give
reliable structures for the outer layers and depths of surface
convection zones. A more thorough investigation of
polluted WDs with helium-dominated atmospheres in \mesa\ awaits
extensions to atmosphere capabilities that can account for these
issues.

\subsubsection{Rotation}
Rotational mixing and its interplay with other fluid processes can be
important in stars \citep{Sengupta18}.
This potential impact is quantified with the Rossby
number ${\rm Ro} = U/2 \Omega L$, where $\Omega$ is the rotational
frequency, and $U$ and $L$ are the characteristic velocity and length
scale for the relevant fluid process.
Large values of the Rossby number indicate that rotation is not
expected to have a strong influence, while ${\rm Ro} \lesssim 1$
indicates potential for significant modifications.
\cite{Sengupta18} studied the effect of rotation on
thermohaline mixing in stellar interiors, where they derived the
Rossby number in an actively mixing region as
\begin{equation}
\label{eq:Rossby}
{\rm Ro} \sim \sqrt{\frac{N^2}{4\Omega^2} \frac{(\varphi/\delta)
    \nabla_\mu}{\nabla_T - \nabla_{\rm ad}}}~,
\end{equation}
where $N$ is the \BV\ frequency. In a non-degenerate WD atmosphere, this
frequency is of order $N^2 \sim g/H$, where ${H = k_{\rm B}T/m_{\rm p}
g}$ is the local pressure scale height. For our polluted WD
models experiencing moderate amounts of thermohaline mixing, we estimate
$(\varphi/\delta)\nabla_\mu/(\nabla_T - \nabla_{\rm ad}) \sim
10^{-4}$ (cf.~\paperone).
We can therefore rewrite Equation~\eqref{eq:Rossby} in terms
of the critical rotation rate $\Omega_{\rm crit} = \sqrt{GM/R^3}$ as
\begin{equation}
{\rm Ro} \sim 10^{-2}\sqrt{\frac R H} \left( \frac{\Omega_{\rm
    crit}}{\Omega}\right)~.
\end{equation}
This requires $\Omega/\Omega_{\rm crit} \gtrsim 10^{-2}$ for rotation
to be important (${\rm Ro} \lesssim 1$). However, typical
isolated WD rotation periods are around one day \citep{Hermes17},
while the critical rotation period is on the order of a few seconds,
so we do not expect rotation to influence the thermohaline mixing in
typical polluted WDs.

Thermohaline mixing has also been discussed as a mechanism for
explaining observed surface abundances in low-mass giant stars
\citep{Charb07,Deniss08,Cantiello10}, but this may require an
implausibly large mixing efficiency $\alpha_{\rm th} > 100$ for
implementations such as Equation~\eqref{eq:KRTD}. This
mixing efficiency appears to be inconsistent with the mixing found in
our models based on \cite{Brown13}.
\cite{Sengupta18} suggested that the interplay of rotation with
thermohaline instability may enhance mixing near the cores of giant
stars. Since we estimate that rotation would be irrelevant for
thermohaline mixing in polluted WDs, this may alleviate the apparent
tension between thermohaline mixing efficiency inferred in these
different regimes.

\begin{figure*} 
\begin{center}
\includegraphics[width=\apjcolwidth]{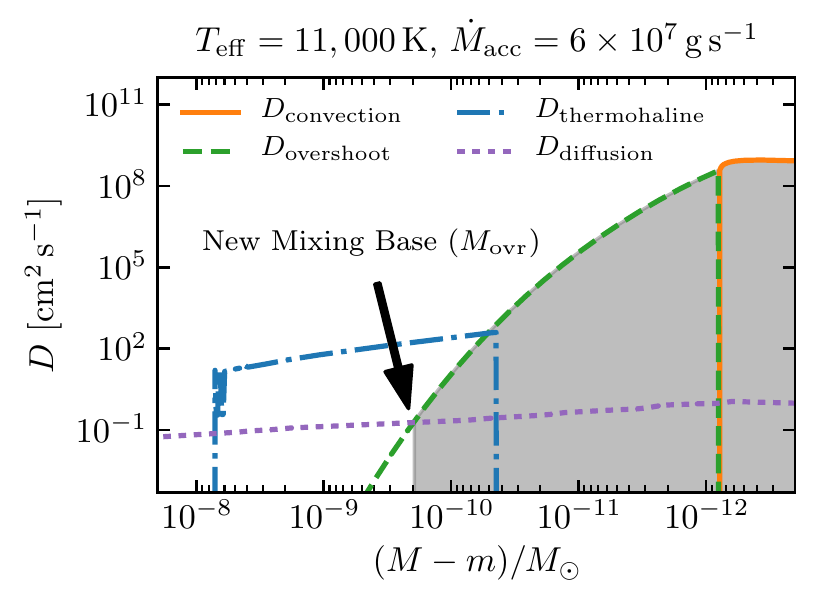}
\hspace{2em}
\includegraphics[width=\apjcolwidth]{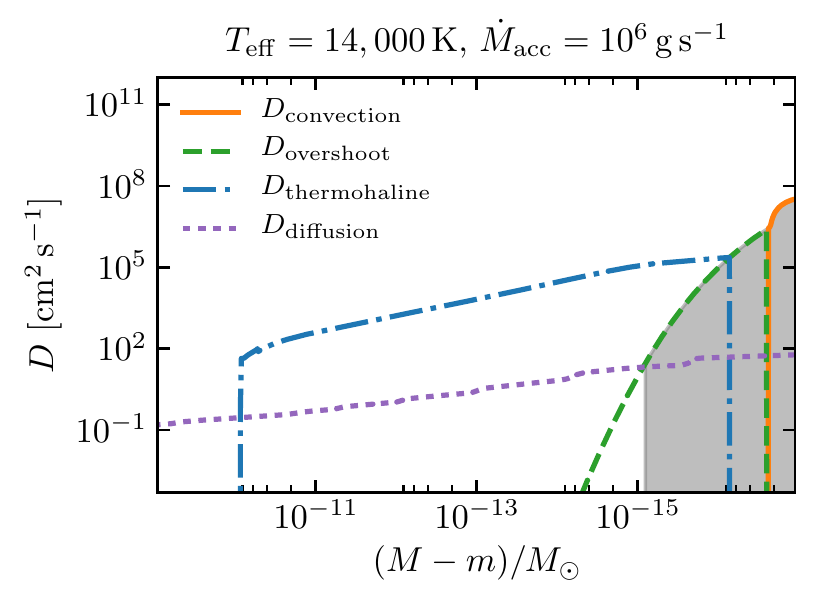}
\caption{
Mixing coefficient profiles for two \mesa\ models. The gray shaded
region represents the extent of surface layers that are expected to be
fully mixed regardless of accretion rate, encompassing a mass of
$M_{\rm ovr}$.
The element diffusion coefficient shown for $D_{\rm diffusion}$ is
that of iron in hydrogen. 
Choosing a different element to represent the particle diffusion
coefficient would result in small changes to the crossover point that
defines $M_{\rm ovr}$, but the steep decline of $D_{\rm overshoot}$
makes these variations negligible.
}
\label{fig:ovr_coeffs}
\end{center}
\end{figure*}

\begin{figure} 
\begin{center}
\includegraphics[width=\apjcolwidth]{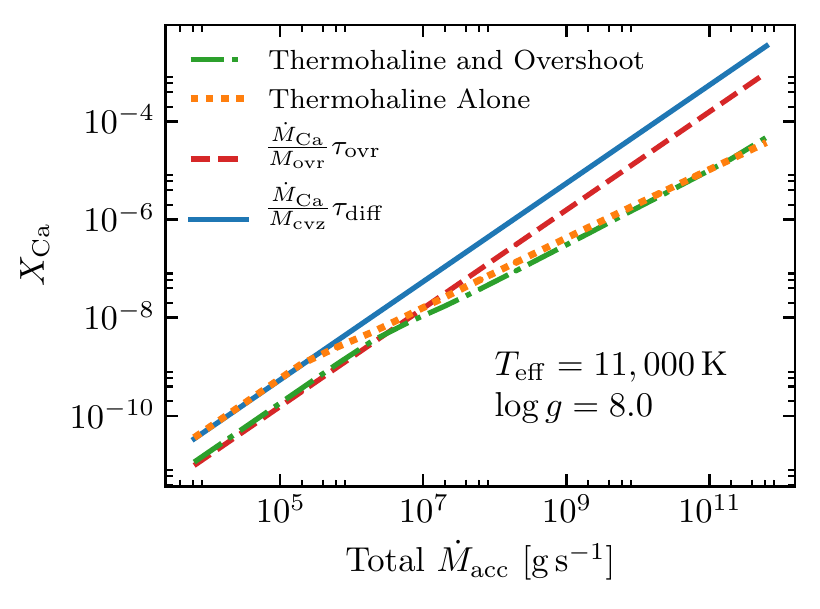}
\caption{
Surface Ca mass fraction after running \mesa\ models for many
diffusion timescales. All models accrete bulk earth composition.
The models for thermohaline alone are the same
as those in Section~\ref{S.Thermohaline}.
The models including overshoot were run for $\approx 10 \, \tau_{\rm
  ovr}$ due to the fact that $\tau_{\rm ovr} \gg \tau_{\rm diff}$.
}
\label{fig:ovr_sep}
\end{center}
\end{figure}

\begin{figure} 
\begin{center}
\includegraphics[width=\apjcolwidth]{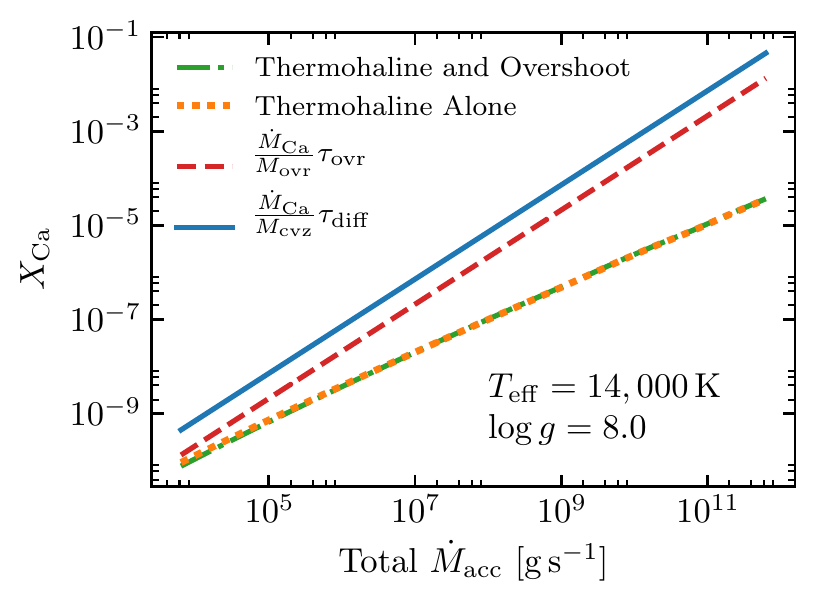}
\caption{
Same as Figure~\ref{fig:ovr_sep} but with a model at $T_{\rm eff} =
14,000 \, \rm K$. Overshoot does not cause changes because significant
thermohaline mixing occurs even at very low accretion rates.
}
\label{fig:ovr_sep14K}
\end{center}
\end{figure}

\subsection{Overshoot}
\label{S.Overshoot}

Convective overshoot may cause fluid motions that can keep
composition thoroughly mixed well below the formal boundary for
convective instability according to the Ledoux criterion
\citep{Freytag96,Koester09,Tremblay15}, even in the
absence of thermohaline instability. This will lead to a larger
effectively mixed region and longer diffusion timescales for a given
$T_{\rm eff}$ \citep{Brassard15,Tremblay17}.

To estimate mixing due to overshoot beneath the convective zone, we
follow the results of \cite{Tremblay15} and
use a diffusion coefficient that decays exponentially with pressure
scale height:
\begin{equation}
\label{eq:overshoot}
D_{\rm overshoot}(r) = D_{0} \exp \left( -
  \frac{2\abs{r-r_0}}{H_P}\right)~,
\end{equation}
where $r_0$ is the radial coordinate of the base of the convection
zone, $H_P$ is the pressure scale height there, and $D_0$ is the mixing
coefficient from MLT near that location.
Figure~\ref{fig:ovr_coeffs} shows the resulting diffusion
coefficient profiles for two \mesa\ models.
In the absence of thermohaline mixing,
this will lead to a new mass of the fully mixed surface region
($M_{\rm ovr}$) defined by the location where the overshoot mixing 
decays to where element diffusion takes over
($D_{\rm overshoot} < D_{\rm diffusion}$, see Figure~\ref{fig:ovr_coeffs}).
There is then a corresponding new diffusion
timescale for each element 
\begin{equation}
\tau_{{\rm ovr},i}\equiv\frac{M_{\rm ovr}}{4\pi r^2\rho v_{{\rm ovr},i}}~,
\end{equation}
where $\rho$, $r$, and $v_{{\rm ovr},i}$ are all evaluated at the base
of the new mixing region defined by $M_{\rm ovr}$. 
The equilibrium observable abundance of an accreted element will then be
\begin{equation}
\label{eq:ovr_eq}
{X_{{\rm eq},i} = \frac{\dot M_i}{M_{\rm ovr}} \tau_{{\rm ovr},i}}
\end{equation}
instead of the analogous value given in Equation~\eqref{eq:tau_eq}.

Figure~\ref{fig:ovr_sep} shows how observable abundances of accreting
metals change when including this form of overshoot in our \mesa\
models. For models at $T_{\rm eff} = 11,000 \, \rm K$, the new
diffusion timescale for Ca is $\tau_{\rm ovr} = 1300 \, \rm yr$,
almost $100$ times larger than the timescale without overshoot. The
larger mixing region means that accreted metals are more diluted for a
given accretion rate, and so larger accretion rates are needed for
thermohaline mixing to cause the observable abundances to diverge from
the prediction of Equation~\eqref{eq:ovr_eq}.
Still, for accretion rates of $\dot
M_{\rm acc} > 10^7 \, \rm g \, s^{-1}$, thermohaline mixing begins to
dominate the final observed abundance, and overshoot causes only small
adjustments when thermohaline mixing is active (see also the left
panel of Figure~\ref{fig:ovr_coeffs}). For the case of a $T_{\rm eff}
= 14,000 \, \rm K$ WD shown in Figure~\ref{fig:ovr_sep14K}, overshoot
extends the small surface mixing region to $M_{\rm ovr} \approx 8
\times 10^{-16} \, M_\odot$ (see right panel in
Figure~\ref{fig:ovr_coeffs}), but this is still so thin that
thermohaline mixing dominates even for modest accretion rates.

While the results shown in this section may serve as a useful
qualitative description of effects that can be expected from
overshoot, 
it is likely that the overshoot mixing prescription given in
Equation~\eqref{eq:overshoot} is too simplistic for WD pollution
applications.
Simulations are beginning to probe regimes specific to convective
overshoot in WDs \citep{Montgomery04,Tremblay15,Kupka18}, and
they appear to show that a simple exponential decay in the diffusion
coefficient is only accurate within a few scale heights of the
convective boundary. Extrapolation down to the much smaller diffusion
coefficients relevant for particle diffusion is likely inaccurate.
Simulations by \cite{Lecoanet16} found overshoot mixing that decays
with a Gaussian profile ($D_{\rm overshoot} \propto
\exp[-(r-r_0)^2/2H^2]$ for some scale height $H$) rather than the
exponential of Equation~\eqref{eq:overshoot}.
A few more recent results appear to confirm this Gaussian overshoot
profile in other contexts \citep{Jones17,Korre18}.
This faster decay of the diffusion coefficient would 
imply that the extra extent of overshoot mixing is smaller
than what is shown in Figure~\ref{fig:ovr_coeffs}.
We therefore refrain from a complete exploration of \mesa\ models
including overshoot until simulations can provide better constraints
on overshoot mixing well below convective boundaries.

Our results are sufficient to conclude that overshoot
will have negligible effects on most accretion rate inferences for
$T_{\rm eff} \gtrsim 12,000 \, \rm K$, where thin surface mixing
regions result in strong concentrations of metals that make
thermohaline mixing dominant. For lower temperatures,
Figure~\ref{fig:ovr_sep} suggests that overshoot may cause significant
adjustments to accretion rate inferences in cases where thermohaline
mixing is not active. Even at higher temperatures, the new
timescales due to overshoot may be important for decay phases where
supply of fresh accreted material has ended and there is nothing to
maintain the $\nabla_\mu$ needed to drive thermohaline instability
near the surface mixing region. In this case, the $M_{\rm ovr}$ and
$\tau_{\rm ovr}$ parameters will govern the exponential decay of
observable surface abundances.

\section{Discussion}
\label{S.discussion}


A significant fraction of WDs show evidence of pollution \citep{Koester14},
and if this fraction represents the fraction of the lifetime of each
individual WD that it is polluted, then Figure~\ref{fig:thermo_mdot}
may be taken as approximately showing a complete history of accretion
rates experienced over a WD lifetime.
In this case, the total mass of planetesimal material
accreted over a WD lifetime would be dominated by the high rates
experienced by young WDs, yielding a high estimate of
${M_{\rm tot} \sim (3 \times 10^8 \, {\rm yr})(10^{12} \, {\rm g \,
  s^{-1}}) \sim 10^{28} \, \rm g}$. However,
\cite{Koester14} point out that the Ca based sample used to
construct Figure~\ref{fig:thermo_mdot} may be biased toward objects
that are especially heavily polluted, since the optical Ca lines used
to select this sample require higher Ca abundances to be detectable
for $T_{\rm eff} \gtrsim 15,000 \, \rm K$ as total WD flux moves
primarily into the UV. While this is unlikely to change the accretion
rates inferred for the objects shown in Figure~\ref{fig:thermo_mdot},
it could hide a much larger intrinsic scatter in the accretion rates
for young DA WDs. Therefore, $10^{28} \, \rm g$ could be an
overestimate of the total mass accreted over a WD lifetime.

This sample of polluted WDs may reveal that some young WDs are
undergoing short timescale bursts of accretion such as those suggested by
\cite{Rafikov11b, Metzger12}. This may help explain the discrepancy
with DBZ WDs for $T_{\rm eff} \gtrsim 15,000 \, \rm K$. The rates here
are too high to be explained by Poynting-Robertson drag
\citep{Rafikov11}, but rare runaway bursts would leave very different
observational signatures for DA and DB WDs \citep{Farihi12}.
DA WDs approach a quasi-equilibrium surface abundance within days or years
in this temperature range even for our \mesa\ models including
thermohaline mixing. On the other hand,
 the diffusion timescales in DB WDs are of order
$10^5$-$10^6$ years, and bursts lasting less than $10^4$ years would
never approach an equilibrium surface pollution level suggesting a
high rate. Instead, DB WD surfaces may represent a more accurate
estimate of accretion rates averaged over their much longer diffusion
timescales.
A more conservative mass estimate for total planetesimal material
may then be  ${M_{\rm tot} \sim (3 \times 10^8 \, {\rm yr})(10^{10} \,
  {\rm g \, s^{-1}}) \sim 10^{26} \, \rm g}$.

Alternative processes could supply polluting material for longer
timescales at rates higher than the limits of Poynting-Robertson drag,
e.g., collisional cascades \citep{Kenyon17a, Kenyon17b} or viscous
evolution of earth-mass dust disks \citep{Lieshout18}. Hence, short
bursts are not strictly necessary to explain the rates shown in
Figure~\ref{fig:thermo_mdot}, but longer timescale processes may
then require that the planetesimal environments form with
significantly different amounts of mass around DA and DB WDs.
\cite{Wyatt14} found that stochastic accretion of a distribution of
planetesimal sizes may be able to explain some discrepancies in
inferred accretion rates for DA and DB WDs without the need to appeal
to large bursts, but this analysis assumed accretion rates inferred
without accounting for thermohaline mixing.

Finally, we note that some authors have pointed out trends
of inferred accretion rates that decline with WD age over timescales of Gyr
\citep[e.g.,][]{Hollands18,Chen18}, consistent with slow depletion of
the planetesimal reservoirs that can obtain highly eccentric orbits on
which they will eventually be tidally disrupted \citep[e.g.,][]{Debes12,Mustill18}.
Our results appear to suggest that this decline may be more dramatic
during the first Gyr of evolution when thermohaline mixing is
accounted for. In particular, the broken power-law for accretion rates
over time used by \cite{Chen18} may not be necessary for rates
inferred using our \mesa\ models. Instead, a single power-law may work
for all WD ages, consistent with the rate at which asteroids
dynamically encounter the WD in the model of \cite{Chen18}.

\section{Conclusions}
\label{S.conclusion}

We have confirmed the result of \paperone\ that thermohaline mixing in
polluted DA WDs with $T_{\rm eff} \gtrsim 10,000 \, \rm K$
requires accretion rates several orders of magnitude
larger than calculations assuming only gravitational sedimentation.
We have provided results from an expanded grid of models to allow
interpolation in $\log g$ as well as $T_{\rm eff}$
\citep[\url{https://doi.org/10.5281/zenodo.2541235},][]{Zenodo18}.
We also find that thermohaline mixing is inconsequential in
polluted DB WDs with $T_{\rm eff} \lesssim 18,000 \, \rm K$ due to
much more massive surface convection zones. Polluted DA WDs
experience a regime of accretion rates low enough that thermohaline
mixing is not active (Table~\ref{tab:critical}), and so
Table~\ref{tab:SMtau} provides diffusion timescales based on our
\mesa\ models.
These timescales are also relevant for WDs where accretion is no
longer ongoing, as they govern the exponential decay of metals sinking
away from the surface where thermohaline mixing is no longer active,
even when it was active during accretion.
Finally, we have also provided a qualitative description of the effects of
convective overshoot, though we refrain from a full exploration of its
effects due to quantitative uncertainty in the overall extent of
overshoot. However, we note that for WDs with thin surface convection
zones ($T_{\rm eff} \gtrsim 12,000 \, \rm K$), thermohaline mixing
dominates down to layers deeper than overshoot can extend, and hence
we do not expect significant modifications to inferred accretion rates
in this regime.

\acknowledgments
{\it Acknowledgments}:
We are grateful to the anonymous referee for comments that improved
the clarity of the paper. We are grateful to 
Matteo Cantiello, Tim Cunningham, Jay Farihi, Gilles Fontaine, Boris
G{\"a}nsicke, Pascale Garaud, Marc Pinsonneault, Josiah Schwab,
Sutirtha Sengupta, Andrew Swan, and Pier-Emmanuel Tremblay
for inspiring and encouraging discussions relating to many topics in
this paper.
We thank Bill Paxton for continuous efforts in support of
broad \MESA\ usage.
This work was supported by the National 
Science Foundation through grants
PHY 17-148958 and ACI 16-63688.

\software{
  \mesa\ \citep{Paxton11,Paxton13,Paxton15,Paxton18},
  \texttt{Astropy} \citep{Astropy13,Astropy18},
  \texttt{Matplotlib} \citep{Matplotlib},
  \texttt{NumPy} \citep{Numpy},
  \texttt{SciPy} \citep{scipy},
  \texttt{MesaScript} \citep{Mesascript}
}

\appendix

\section{Diffusion Coefficients for Neutral Atoms}
\label{S.neutral}

When diffusion occurs near the surface of a star, even a very small
fraction of neutral particles for a given species can have a dramatic
effect on the net diffusion flux for the element.
The Coulomb collision formalism for diffusion coefficients no longer
applies for neutral atoms, and induced dipole scattering of the
neutral atoms with background ions becomes the relevant physical
process for diffusion \citep{Vennes11proc,Vennes11}.
\mesa\ does not currently offer options
for diffusion coefficients based on dipole scattering, but Section~9
of \cite{Paxton15} describes much of the formalism necessary to
construct them.
Referring to Equations (84)--(86) of \cite{Paxton15}, we see that
the relevant coefficients for the \cite{Burgers69} diffusion
equations are expressed in terms of
\begin{equation}
\Sigma_{st}^{(lj)} = \frac{4 \pi}{\pi^{3/2}} \int_0^\infty dv \, \exp
\left( \frac{-v^2}{\alpha_{st}^2} \right)
\frac{v^{2j+3}}{\alpha_{st}^{2j+4}} S_{st}^{(l)}~,
\end{equation}
where $\alpha_{st} = 2 k_BT/\mu_{st}$, $\mu_{st} = m_s m_t/(m_s +
m_t)$ is the reduced mass of the scattering particles, and
\begin{equation}
S_{st}^{(l)} = 2 \pi \int_0^\infty(1-\cos^l \chi_{st}) b \, db
\end{equation}
are the traditional scattering cross section integrals. The scattering
angle $\chi_{st}$ is a function of both impact parameter $b$ and
relative velocity $v$, depending on the physics of the scattering
process between particles $s$ and $t$. For dipole scattering,
Chapter~2 in \cite{Draine11} gives the scattering cross section of an
ion with a polarizable neutral atom as
\begin{equation}
S_{st}^{(1)} = 2.41 \pi Z e \left( \frac{\alpha_N}{\mu_{st}} \right)^{1/2} \frac 1 v~,
\end{equation}
where $Z$ is the charge of the ion and $\alpha_N$ is the
polarizability of the neutral atom.
The result for $\Sigma^{(11)}_{st}$ is then
\begin{equation}
\Sigma_{st}^{(11)} 
= 3.62 \pi \frac{Ze}{\alpha_{st}} \left( \frac{\alpha_N}{\mu_{st}} \right)^{1/2}~.
\end{equation}
Hence the result for the resistance coefficient for ions scattering
with induced dipoles of neutral atoms is
\begin{equation}
\label{eq:neutralK}
K_{st} = \frac 2 3 n_s n_t \mu_{st} \alpha_{st} \Sigma_{st}^{(11)}
= 2.41 \pi n_s n_t Ze \left( \mu_{st} \alpha_N \right)^{1/2}~.
\end{equation}
For comparison, the resistance coefficient given by \cite{Burgers69}
for diffusion of ions with other ions is 
\begin{equation}
K_{st} = \frac{16 \sqrt \pi}{3} \frac{n_s n_t Z_s^2Z_t^2 e^4}{\mu_{st}
  \alpha_{st}^3} \ln \Lambda~,
\end{equation}
where $\ln \Lambda$ is the Coulomb logarithm.
Diffusion velocities scale inversely with the resistance coefficients
$K_{st}$, so we can use these expressions to estimate the relative
speeds of neutral and ionized metals.

Consider the case of a mixture of singly ionized oxygen and neutral oxygen diffusing in an ionized Hydrogen background. 
The above equations give that the ratio of the ion-neutral coefficient to the ion-ion coefficient is
\begin{equation}
\frac{K_{st}^{(\rm dipole)}}{K_{st}^{(\rm coulomb)}} 
= \frac{3.62 (2 k_B T)^{3/2} \left( \pi \alpha_N \right)^{1/2}}{8
  e^3 \ln \Lambda}~.
\end{equation}
The polarizability of neutral oxygen is given in Table~2.1 of
\cite{Draine11} as $\alpha_N = 5.326 a_0^3$, where $a_0$ is the Bohr
radius. For $T = 15,000 \, \rm K$, we find that the resistance
coefficient ratio is
\begin{equation}
\frac{K_{st}^{(\rm dipole)}}{K_{st}^{(\rm coulomb)}} =
\frac{0.054}{\ln \Lambda}~.
\end{equation}
For conditions near the surface of a WD, we estimate the Coulomb logarithm as
$\ln \Lambda \approx 5$ \citep[][Table~5.1]{Spitzer62}.
The final result for the ratio of coefficients is
$K_{st}^{(\rm dipole)}/K_{st}^{(\rm coulomb)} \sim 10^{-2}$.
This means that the neutral oxygen particles will have diffusion velocities
approximately $100$ times faster than the singly ionized particles. A
mere $1\%$ of particles being neutral for a particular element can therefore
significantly modify the net diffusion flux for that element.
This effect would be most noticeable for metals with relatively high
first ionization potentials.

Since the physics of scattering is fundamentally different for charged and
neutral particles, the diffusion and resistance coefficients do not
scale in a meaningful way as $Z \to 0$. 
It is hence meaningless to adopt an average charge for an
element for purposes of diffusion calculations in the case of
$\bar Z < 1$. This is why the diffusion implementation in \mesa\
currently assumes that all diffusing particles are at least singly
ionized. Extending the implementation to account for neutral particles
would require two improvements:
a) the ability to separate neutral particles off as distinct diffusion
classes, and
b) incorporating tables of atomic polarizabilities to use with
Equation~\eqref{eq:neutralK} for the resistance coefficients to use in
the Burgers equations.

\bibliographystyle{yahapj}

\end{document}